\documentclass[useAMS,usenatbib]{mn2e}
\newcommand{\mnras}{MNRAS}
\newcommand{\aap}{A\&A}
\newcommand{\apj}{ApJ}
\newcommand{\aj}{AJ}
\usepackage{txfonts}
\usepackage{graphicx}
\title[Aberration in proper motions for Galactic stars]{Aberration in proper motions for stars in our Galaxy}
\author[J.-C. Liu, Y. Xie and Z. Zhu]{J.-C. Liu,\thanks{E-mail:
jcliu@nju.edu.cn} Y. Xie and Z. Zhu\\
School of Astronomy and Space Science, Nanjing University, 210093 Nanjing, PR China\\
Key Laboratory of Modern Astronomy and Astrophysics, Ministry of Education (Nanjing University), 210093 Nanjing, PR China}
\begin{document}

\date{Accepted 2013 June 05. Received 2013 June 01; in original form 2013 April 11}

\pagerange{\pageref{firstpage}--\pageref{lastpage}} \pubyear{2013}

\maketitle

\label{firstpage}

\begin{abstract}
Accelerations of both the solar system barycenter (SSB) and stars in the Milky Way cause a systematic observational effect on the stellar proper motions, which was first studied in the early 1990s and developed by J. Kovalevsky (aberration in proper motions, 2003, \aap, 404, 743). This paper intends to extend that work and aims to estimate the magnitude and significance of the aberration in proper motions of stars, especially in the region near the Galactic center. We adopt two models for the Galactic rotation curve to evaluate the aberrational effect on the Galactic plane. Based on the theoretical developments, we show that the effect of aberration in proper motions depends on the galactocentric distance of stars; it is dominated by the acceleration of stars in the central region of the Galaxy. Within 200 pc from the Galactic center, the systematic proper motion can reach an amplitude larger than 1000 $\rm \mu as\,yr^{-1}$ by applying a flat rotation curve. With a more realistic rotation curve which is linearly rising in the core region of the Galaxy, the aberrational proper motions are limited up to about 150 $\rm \mu as\,yr^{-1}$. Then we investigate the applicability of the theoretical expressions concerning the aberrational proper motions, especially for those stars with short period orbits. If the orbital period of stars is only a fraction of the light time from the star to the SSB, the expression proposed by Kovalevsky is not appropriate. With a more suitable formulation, we found that the aberration has no effect on the determination of the stellar orbits on the celestial sphere. The aberrational effect under consideration is small but not negligible with high-accurate astrometry in the future, particularly in constructing the Gaia celestial reference system realized by Galactic stars.
\end{abstract}


\section{Introduction}
It is well known that the velocity of an observer results in aberration in positions. For example, the Earth's orbital velocity is about 30 $\rm km\, s^{-1}$, which is a fraction $10^{-4}$ of the speed of light in a vacuum. The order of corresponding aberrational displacement is $10^{-4}$ radians, or about $20''$. In a higher hierarchy of reference system \citep{klioner98}, the velocity of the solar system barycenter (SSB) is responsible for the first order aberration in position of about $150''$, however this value is a constant and not detectable. In addition, the acceleration of an observer produce aberrational effect in proper motions of celestial objects (\citealt{kovalevsky03}, \citealt{kopeikin06}), which is a variational effect with respect to the aberration in position. Given astrometric measurements at micro-arcsecond level, the aberration in proper motions resulting from the acceleration of the SSB has impact on the celestial reference system realized by extragalactic radio sources (ICRS) and Earth rotation parameters, thus should be considered in the near future (\citealt{titov10}; \citealt{liu12}). Using VLBI observations over more than 20 years, this effect has been detected by \citet{titov11}.

Let us first recall the effect of the planetary aberration, which depends on the relative velocity of the Earth and a planet in the solar system:
\begin{equation}
\label{eq_pl}
\Delta \bmath p^{\rm PL} = \frac{1}{c}\bmath p\times [(\bmath v^{\rm E} - \bmath v^{\rm PL}) \times \bmath p].
\end{equation}
where $\bmath p$ is the true geometric position vector, $\bmath v^{\rm E}$ and $\bmath v^{\rm PL}$ the velocities of the Earth and the planet under consideration, and $\Delta \bmath p^{\rm PL}$ the change of direction. Because the position and velocity of planets and the Earth are known accurately in the solar system, this planetary aberration can be fully corrected.

For stars in the Milky Way and an observer at the SSB, the aberration in positions coming from the velocity of the SSB and star is similar to Eq. (\ref{eq_pl}):
\begin{equation}
\label{eq_star}
\Delta \bmath p^{\rm S} = \frac{1}{c}\bmath p\times [(\bmath v^{\rm B} - \bmath v^{\rm S}) \times \bmath p],
\end{equation}
where the superscript `B' and `S' represent the SSB and star, respectively.
The aberration in proper motions for stars can be written as the first time derivative of the above equation:
\begin{equation}
\label{eq_vec}
\Delta \dot {\bmath p}^{\rm S} = \frac{1}{c}\bmath p\times [(\bmath a^{\rm B} - \bmath a^{\rm S}) \times \bmath p].
\end{equation}
The proper motion (independent of distance of stars) resulting from $\bmath a^{\rm B}$ is the same as the effect for extragalactic sources, which forms a dipolar field on the celestial sphere from the anti-Galactic center to the Galactic center. The second part corresponding to the acceleration $\bmath a^{\rm S}$ is more complicated and will be discussed in detail in the following. Considering that the magnitude of accelerations of stars are comparable or even larger than that of the SSB, both accelerations give rise to aberrational proper motions for stars.

This paper is in part an extension and improvement of \citet{kovalevsky03} for its 7th section where only compendious description was presented regarding `time-dependent aberration of stars'. In Section 2, theoretical expressions for the combined aberration in proper motions defined by two kinds of rotation curves of the Galaxy are developed. As the effect is remarkably larger near the Galactic center, we focus our analysis within this area in Section 3 and provide necessary remarks on this effect. Then in Section 4, we discuss the influence of the effect of the aberration in proper motions on the future Gaia reference system. Finally, summary and discussion are given in Section 5. Appendix A is a computation applied to the S0-2 star around the central black hole in the Galactic center, for the purpose of interpreting the proper use of theoretical expressions.

\paragraph*{Nomenclature issue}
The aberrational effect on the extragalactic sources attribute to the acceleration of the SSB has been called the `secular aberration' \citep{kopeikin06}, `secular aberration drift' (\citealt{titov10}; \citealt{titov11}; \citealt{xu13}), `Galactic aberration' (\citealt{malkin11}; \citealt{liu12}), and also a more concise term `glide' \citep{mignard12}, which is as easy to be used as `rotation'. Since the aberration in stellar proper motions in the Galaxy depends on both of the acceleration of the SSB and stars themselves, the above terms are not sufficient to distinguish these two phenomena, although they are quite similar, thus a proper terminology is necessary. Following \citet{kovalevsky03}, the nomenclature ``aberration in proper motions'' chosen here is a general one which expresses the observational effect, regardless of the origin.

\section{Theoretical development for aberration in proper motions}
\subsection{Coordinate systems}
We denote the equatorial coordinate system $(X,\,Y,\,Z)$ and the Galactic coordinate system $(X_{\rm G},\,Y_{\rm G},\,Z_{\rm G})$, such that the $X_{\rm G}$ axis is directed to the Galactic center and the $X$-$Y$ plane coincide with the Galactic plane (\citealt{blaauw60}; \citealt{liu11a}). Taking the compact radio source Sgr A* (super massive black hole) as the Galactic center, the equatorial-to-Galactic coordinate transformation matrix should be $\mathcal N_{\rm 2MASS}$ given by Equation (18) of \citet{liu11b}. We also introduce the local tangential Cartesian coordinate system $(x,\,y,\,z)$. At a point $(\alpha,\,\delta)$ in the equatorial coordinate system, the unit vectors in the right ascension, declination, and radial directions are denoted as $(\bmath e_\alpha,\,\bmath e_\delta,\,\bmath e_r)$, while in the Galactic coordinate system at the point $(\ell,\, b)$, the unit vectors in the Galactic longitude, latitude, and radial directions are $(\bmath e_\ell,\,\bmath e_b,\,\bmath e_r)$. For the visual plot of these coordinate systems, readers are refer to Figure 1 of \citet{mignard12}.

The relations between the unit vectors of the triads of $(X,\,Y,\,Z)$ and $(x,\,y,\,z)$ are such that \citep{green85}
\begin{eqnarray}
\label{eq_ead}
\bmath e_\alpha & = & -\sin\alpha\cdot \bmath e_X + \cos\alpha\cdot \bmath e_Y \nonumber \\
\bmath e_\delta & = & - \cos \alpha \sin \delta\cdot \bmath e_X - \sin \alpha \sin \delta\cdot \bmath e_Y + \cos \delta\cdot\bmath e_Z \\
\bmath e_r & = & + \cos\alpha\cos \delta\cdot \bmath e_X+ \sin \alpha \cos \delta\cdot \bmath e_Y + \sin \delta\cdot\bmath e_Z. \nonumber
\end{eqnarray}
For the Galactic coordinate system, we have
\begin{eqnarray}
\label{eq_elb}
\bmath e_\ell & = & -\sin\ell\cdot \bmath e_{X_{\rm G}} + \cos\ell\cdot \bmath e_{Y_{\rm G}} \nonumber \\
\bmath e_b & = & - \cos \ell \sin b\cdot \bmath e_{X_{\rm G}} - \sin \ell \sin b\cdot \bmath e_{Y_{\rm G}} + \cos b\cdot\bmath e_{Z_{\rm G}} \\
\bmath e_r & = & + \cos\ell\cos b\cdot \bmath e_{X_{\rm G}} + \sin \ell \cos b\cdot \bmath e_{Y_{\rm G}} + \sin b\cdot\bmath e_{Z_{\rm G}}. \nonumber
\end{eqnarray}

\subsection{Aberration in proper motions for Galactic stars}
Projecting Eq.~(\ref{eq_vec}) to the local tangential coordinate system $(x,\,y,\,z)$ in the Galactic coordinate system, the aberrational proper motions in longitude and latitude directions can be derived as follows:
\begin{eqnarray}
\label{eq_mulb1}
\Delta \mu_\ell\cos b & = & \Delta \bmath {\dot p}^{\rm S}\cdot \bmath e_\ell = \frac{1}{c}\bmath p\times [(\bmath a^{\rm B} - \bmath a^{\rm S}) \times \bmath p] \cdot \bmath e_\ell \nonumber \\
\Delta \mu_b & = & \Delta \bmath {\dot p}^{\rm S}\cdot \bmath e_b = \frac{1}{c}\bmath p\times [(\bmath a^{\rm B} - \bmath a^{\rm S}) \times \bmath p] \cdot \bmath e_b,
\end{eqnarray}
where $\bmath e_\ell$, $\bmath e_b$ are given in Eq. (\ref{eq_elb}). The third component for the change of the radial velocity is
\begin{equation}
\Delta v_r = \Delta \bmath {\dot p}^{\rm S}\cdot \bmath e_r =\Delta \bmath {\dot p}^{\rm S}\cdot \bmath p = 0,
\end{equation}
because the aberration only change the direction (say, unit vectors) of stars.

\begin{figure}
\label{fig1}
\centering
\includegraphics[width=60mm]{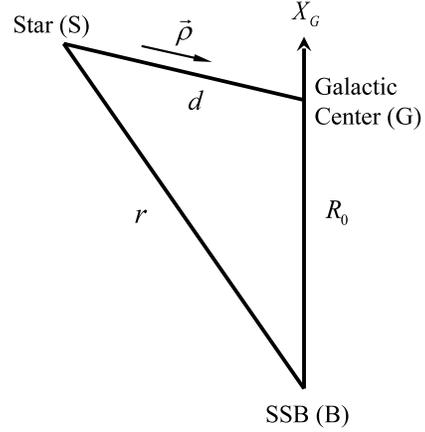}
\caption{The geometrical relation of the SSB, Galactic center, and star. $\vec \rho$ is the unit vector from the star to the Galactic center.}
\end{figure}

In our Galaxy, we suppose that all stars accelerate toward the Galactic center due to the gravitational attraction. In the present study, we assume that all stellar orbits are circular, although there may exists non-circular motions for stars in our Galaxy. The geometrical relation is shown in Fig.~1 for the solar system barycenter (SSB, denoted B), the Galactic center (denoted G), and a star (denoted S). The star is located at $(\ell, b, r)$, where $r$ is the the distance to the SSB. The unit vector of the acceleration of the star is denoted $\bmath \rho$ in the plot. The vector $\bmath {BS}$ and $\bmath {BG}$ can be written as
\begin{eqnarray}
\bmath {BS} &=& \left(r\cos\ell\cos b,\, r\sin\ell\cos b,\, r\sin b\right)^{\rm T}\nonumber \\
\bmath {BG} &=& \left(R_0,\, 0,\,0\right)^{\rm T},
\end{eqnarray}
in which $R_0\simeq8.0\,\rm kpc$ \citep{reid93} is adopted as the distance between the SSB and the Galactic center.
The distance of the star from the Galactic center (denoted $d$) is then calculated from $\bigtriangleup BGS$:
\begin{equation}
d = \sqrt{R_0^2 + r^2 -2R_0r\cos\ell\cos b},
\end{equation}
and the unit vector in the direction of $\bmath{SG}$ can be derived as
\begin{eqnarray}
\label{eq_rho}
\bmath \rho & = & \frac{\bmath{SG}}{d} = \frac{1}{d}\left(R_0 - r\cos\ell\cos b,\, -r\sin\ell\cos b,\, -r\sin b\right)^{\rm T}\nonumber \\
& \equiv & \left(\rho_1,\,\rho_2,\,\rho_3\right)^{\rm T}.
\end{eqnarray}
Therefore the acceleration of the SSB and star are respectively
\begin{eqnarray}
\bmath a^{\rm B} &=& a^{\rm B} \cdot \bmath e_{X_{\rm G}} \nonumber \\
\bmath a^{\rm S} &=& a^{\rm S} \cdot \bmath \rho.
\end{eqnarray}
We define following parameters in the unit of proper motions:
\begin{eqnarray}
\label{eq_defAB}
A^{\rm B} & = & \frac{a^{\rm B}}{c} = \frac{V_0^2}{cR_0}\simeq 5\,{\rm \mu as\, yr^{-1}}\nonumber \\
A^{\rm S} & = & \frac{a^{\rm S}}{c} = \frac{a^{\rm B}}{c} \frac{a^{\rm S}}{a^{\rm B}} =A^{\rm B}\frac{a^{\rm S}}{a^{\rm B}},
\end{eqnarray}
where the former quantity $A^{\rm B}$ (constant) corresponding to the SSB was called the `Galactic aberration constant' in \citet{malkin11} and \citet{liu12}. The magnitude of $A^{\rm B}$ is about $4-6~\rm\mu as~yr^{-1}$, based on the recent determination of Galactic constants (e.g. \citealt{bobylev10}; \citealt{reid09}; \citealt{zhu09}) and its value in Eq. (\ref{eq_defAB}) is chosen to match with our previous study \citep{liu12} \footnote{The constant $A^{\rm B}$ is closely related to the Galactic constants $V_0$ and $R_0$. $A^{\rm B} = 5\,{\rm \mu as\, yr^{-1}}$ (adopted value in this paper) corresponds to $V_0 = 240\,{\rm km\,s^{-1}}$ and $R_0 = 8.0\,{\rm kpc}$, while $A^{\rm B} = 4\,{\rm \mu as\, yr^{-1}}$ corresponds to $V_0 = 220\,{\rm km\,s^{-1}}$ and $R_0 = 8.5\,{\rm kpc}$, as adopted by \citet{kovalevsky03}.}. Then we define a parameter $\gamma$ as the ratio of the accelerations of the star and SSB:
\begin{equation}
\label{eq_gamma}
\gamma = \frac{a^{\rm S}}{a^{\rm B}}.
\end{equation}
Using the above formulas and definitions, the proper motions in Eq. (\ref{eq_mulb1}) can be written as:
\begin{eqnarray}
\label{eq_mulb2}
\Delta \mu_\ell\cos b & = &  A^{\rm B}\bmath p\times [(\bmath e_{X_{\rm G}} - \gamma \bmath \rho) \times \bmath p] \cdot \bmath e_\ell \nonumber \\
\Delta \mu_b & = &  A^{\rm B} \bmath p\times [(\bmath e_{X_{\rm G}} - \gamma \bmath \rho) \times \bmath p] \cdot \bmath e_b,
\end{eqnarray}
Expanding Eq. (\ref{eq_mulb2}), we obtain
\begin{eqnarray}
\label{eq_mulb3}
\Delta \mu_\ell\cos b & = & A^{\rm B}[
-\left(1-\gamma\rho_1\right)\sin\ell-\gamma\rho_2\cos\ell] \nonumber \\
& = & - A^{B}\left(1-\gamma \frac{R_0}{d}\right)\sin\ell \nonumber \\
\Delta \mu_b & = & A^{\rm B}[
-\left(1-\gamma\rho_1\right)\cos\ell\sin b
+\gamma\rho_2 \sin\ell \sin b \nonumber \\
&& -  \gamma\rho_3 \cos b] \nonumber \\
& = & - A^{B}\left(1-\gamma \frac{R_0}{d}\right)\cos\ell\sin b,
\end{eqnarray}
in which $\rho_1$, $\rho_2$, and $\rho_3$ are the three components of the unit vector $\bmath \rho$ as defined in Eq. (\ref{eq_rho}). For extragalactic radio sources whose accelerations are zero or too small to be detected (i.e. $\bmath a^{\rm S} = 0$, $\gamma=0$), Eq. (\ref{eq_mulb3}) degenerates into the form of the pure dipolar proper motion field (see, e.g. Eq. (4) of \citealt{liu12}).

\subsection{Aberration in proper motions based on rotation curves of the Galaxy}
In order to evaluate the magnitude of aberrational proper motions, it is necessary to know the accelerations of stars, or equivalently the parameter $\gamma$ in (\ref{eq_gamma}). Since accelerations of stars are not available one by one, certain statistical models for the Galactic kinematics, such as rotation curves, are necessarily be used. For stars in the Galaxy, especially for those on the Galactic disk, every star revolves more or less around the Galactic center. One simplified case is such that the rotation curve is completely flat throughout the Galaxy (see Fig. 2(a), which is also the assumption adopted by \citet{kovalevsky03} in his calculation). In this case, the rotation velocity of a star is the same as the velocity of the SSB:
\begin{equation}
V^{\rm S} = V_0,
\end{equation}
and the acceleration in the circular orbit is
\begin{equation}
a^{\rm S} = \frac{\left(V^{\rm S}\right)^2}{d} = \frac{V_0^2}{d} = \frac{V_0^2}{R_0}\frac{R_0}{d},
\end{equation}
so that
\begin{eqnarray}
\label{eq_gamma}
\gamma = R_0/d.
\end{eqnarray}
Inserting (\ref{eq_gamma}) into Eq. (\ref{eq_mulb3}), one has the resulting proper motions in Galactic longitude and latitude:
\begin{eqnarray}
\label{eq_mul}
\Delta \mu_\ell\cos b & = & -A^{\rm B}\left[1-\left(\frac{R_0}{d}\right)^2\right]\sin\ell  \\
\label{eq_mub}
\Delta \mu_b & = & -A^{\rm B}\left[1-\left(\frac{R_0}{d}\right)^2\right]\cos\ell \sin b.
\end{eqnarray}

\begin{figure}
\label{fig2}
\centering
\includegraphics[width=70mm]{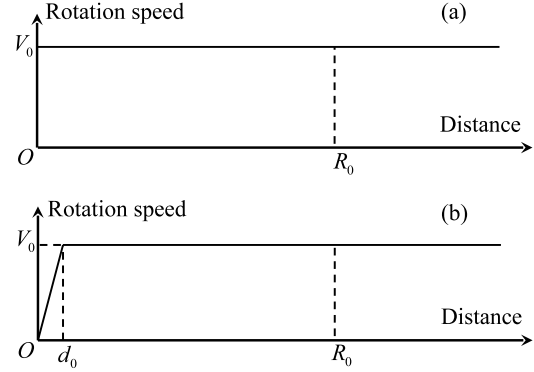}
\caption{The simplified rotation curves of the Galaxy. The origin of both plots is the Galactic center. (a) Flat rotation curve at $V_0$; (b) linear rotation curve up to $d_0$ and then constant from $d_0$ at $V_0$.}
\end{figure}

In the brackets of the above expressions, the first terms with magnitude $A^{\rm B}$ are caused by the acceleration of the SSB, which construct a distance-independent dipole proper motion field. The second terms are inversely proportional to the squared distance of the star to the Galactic center. This means that the proper motions increase as the stars are closer to the Galactic center. On the other hand, the terms $\sin \ell$ and $\sin b$ decrease as the star goes closer to the Galactic center. The final magnitudes of proper motions are mutually affected by these two factors. When applying rotation curves, we can only calculate the aberration for stars on (or nearby) the Galactic plane, because stars belong to other components of the Galaxy (e.g. halo) do not follow the property of the rotation curve. Figure 3 shows the contour plot for $\Delta \mu_\ell\cos b$ (note that $\Delta \mu_b=0$ since $b=0$). The upper panel is for a wilder range up to 12 kpc in $X$ and $Y$ directions centered on the Galactic center, and the lower plot is enhancement around the center up to 300 pc. Note that Fig.~3 is consistent with Figure 5 and 6 of \citet{kovalevsky03}, and that the effect of aberration disappears on the $X_{\rm G}$ axis where $\ell = b = 0$ and on the Galactocentric ring passing by the SSB where $d = R_0$. In the area far away from the Galactic center, the aberrational effect is raising from both of the acceleration of the SSB and the star, while in the central region, this effect is dominated by the acceleration of stars with large values for the parameter $\gamma$. Refer to the bottom plot, a remarkable aberrational proper motion of $1000\,\rm \mu as\,yr^{-1}$ is achieved at about 200 pc apart from the Galactic center.

\begin{figure}
\label{fig3}
\centering
\includegraphics[width=80mm]{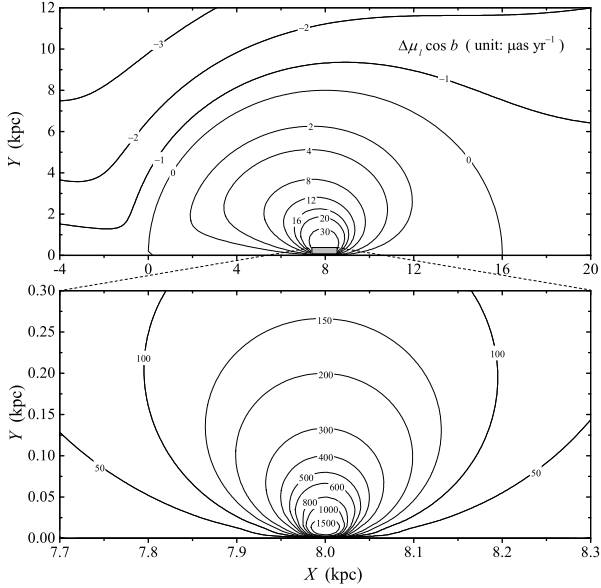}
\caption{The amplitude of aberration in proper motions on the Galactic plane ($b=0$, Eq. \ref{eq_mul}) corresponding to the flat rotation curve in Fig. 2(a). The Galactic center is located at $(8.0,\,0)$ and the SSB is at the $(0,\,0)$ point. The upper plot covers almost the whole Galactic plane, while the lower plot is the enlarged drawing for the vicinity of the Galactic center. Note that in this plot (idem for Figs. 4 and 5) $X$ and $Y$ are measured along the axes of $X_{\rm G}$ and $Y_{\rm G}$ in the Galactic coordinate system; they should be distinguished from the axes in the equatorial coordinate system as defined in Sect. 2.1}
\end{figure}

Describing the rotation curve with only flat straight line is not appropriate for the inner part the Galaxy, since the rotation of the bulge is nearly rigid. A more reasonable approximation of the rotation curve is the one as shown in Fig. 2(b), which separate the bulge from the disk at the boundary $d=d_0$. Out of the bulge, the rotation curve is almost flat due to various studies (e.g. \citealt{fich89}; \citealt{frink96}; \citealt{pont97}). The inner truncation length of the bulge of the Galaxy is only about $100-300$ pc (\citealt{binney97}; \citealt{bissantz02}; \citealt{vanholl09}). The expression of the rigid rotation speed of stars within the boundary $d_0$ is
\begin{equation}
\label{eq_vs}
V^{\rm S} = V_0\frac{d}{d_0}~~~~~~~~(d<d_0),
\end{equation}
and the parameter $\gamma$ in this region is
\begin{equation}
\label{eq_gamma2}
\gamma = \left(\frac{V^{\rm S}}{V_0}\right)^2\frac{R_0}{d} = \frac{R_0d}{d_0^2}.
\end{equation}
It is proportional to the galactocentric distance of the star ($d$), while in the previous case it is in inverse proportion to $d$. Inserting Eq. (\ref{eq_gamma2}) into Eq. (\ref{eq_mulb3}), we obtain the resulting aberrational proper motions in Galactic longitude and latitude:
\begin{eqnarray}
\label{eq_mul2}
\Delta \mu_\ell\cos b & = & -A^{\rm B}\left[1-\left(\frac{R_0}{d_0}\right)^2\right]\sin\ell  \\
\label{eq_mub2}
\Delta \mu_b & = & -A^{\rm B}\left[1-\left(\frac{R_0}{d_0}\right)^2\right]\cos\ell \sin b~~(d<d_0).
\end{eqnarray}

The term $1-(R_0/d_0)^2$ in brackets is a constant of an order of 1000 (if $R_0=8.0$ kpc and $d_0=0.3$ kpc), consequently the proper motions are independent of distance of stars in the rigid rotation mode. Within the small area of celestial sphere near the Galactic center, the proper motion field is simply a stronger dipole, but due to the fact that both $\sin\ell$ and $\sin b$ are small quantities, the proper motions are not very significant. Out of the bulge, the expressions are the same as Eqs. (\ref{eq_mul}) and (\ref{eq_mub}). Adopting $d_0=0.3\,\rm kpc$, the magnitude of proper motions are shown in Fig. 4, where the contours in the range $d<d_0$ correspond to straight lines of constant Galactic longitudes. In this area, the largest proper motion is only about $150\,\rm \mu as\,yr^{-1}$ at $X = 8.0\,\rm{kpc}$ and $Y = d_0$. Using this kind of rotation curve, the magnitudes of the aberrational proper motions are limited, and this is different from the previous case, for which the aberrational effect can be infinity near the Galactic center.

For recent practical models of the rotation curve of the Milky Way, readers are refer to e.g. \citet{foster10}, \citet{bovy12}, or \citet{ruiz12}. It is also necessary to mention that our calculation in this section is only valid for the stars whose accelerations are directing to the Galactic center. If the reality is not that case, one should apply Eq. (\ref{eq_mulb1}) for a more general discussion.

\begin{figure}
\label{fig4}
\centering
\includegraphics[width=80mm]{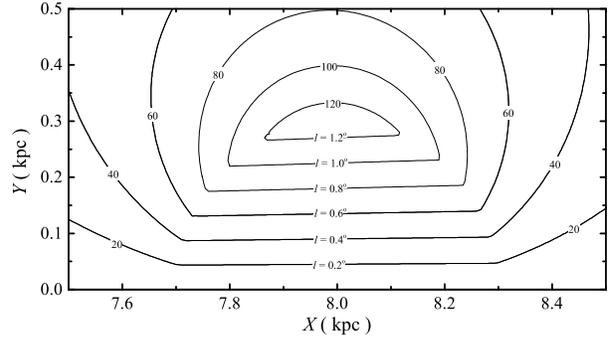}
\caption{The amplitude of aberration in proper motions on the Galactic plane ($\ell=0$, Eq. \ref{eq_mul2}) corresponding to the flat rotation curve in Fig. 2(b). The Galactic center is located at $(8.0,\,0)$ and the SSB is at $(0,\,0)$.}
\end{figure}

\section{Proper use of the expressions for the aberrational effect}
Conceptually, the aberration in proper motions resulting from the stellar acceleration $\bmath a^{\rm S}$ is the variation of projected velocity on the celestial sphere during the light time from the star to the observer (SSB). Written in Eq. (\ref{eq_vec}), the acceleration of the star $\bmath a^{\rm S}$ should be a constant vector, which means that the motion of the star must be (or approximately) rectilinear during the time span of light travel. To this end, we note that Eq. (\ref{eq_vec}) and follow-up expressions that describe the spurious proper motions are simplified, and should be used with caution, especially in the Galactic center region.  An example of improper application of Eq. (\ref{eq_vec}) to a specific star S0-2 at the central region of the Galaxy is presented in appendix A.

\subsection{Working with simplified rotation curve of the Galaxy}
To apply correctly the effect of aberration in proper motions for stars in the galaxy, comparison of the light time from the star to the SSB (denoted $T_{\rm L}$) with the stellar orbit period (denoted $P$) is necessary. With an assumption of circular motion of stars on the Galactic disk, the ratio of the orbital period and the light time is such that
\begin{equation}
\label{eq_tl}
\tau = \frac{P}{T_{\rm L}} = 2\pi\left(\frac{d}{r}\right)\left(\frac{c}{V^{\rm S}}\right),
\end{equation}
where all the symbols are the same as in the previous section. The parameter $\tau$ can be used to represent the degree of linearity of the stellar motion during the time $T_{\rm L}$. If $\tau$ is much larger than 1 (e.g. 100), one can assume safely that the motion of the star is almost linear, or in other words, that the acceleration of the star is approximately a constant ($\tau=+\infty$ means that $\bmath a^{\rm S}$ is rigorously a constant vector). If the rotation curve is flat ($V^{\rm S}=V_0$) as shown in Fig. 2(a), the galactocentric distance of a star should satisfy $d>0.1~\rm{kpc}$ to fulfill  $\tau>100$. The distribution of $\tau$ is plotted in Fig.~5 with an application of flat rotation curve.
\begin{figure}
\label{fig_5}
\centering
\includegraphics[width=80mm]{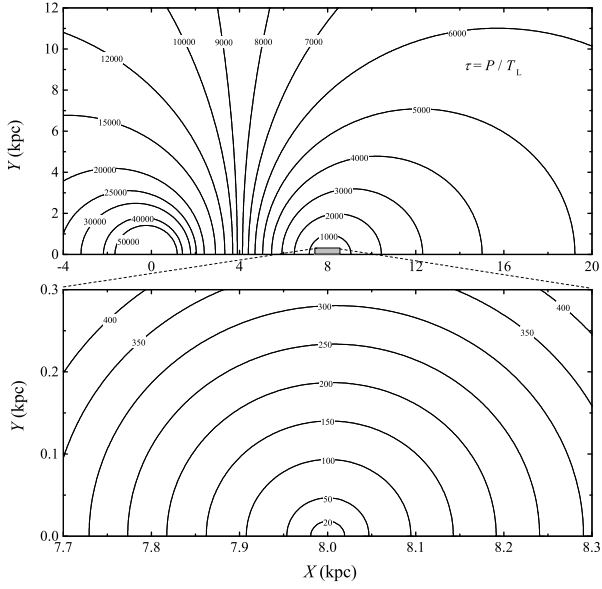}
\caption{Distribution of the parameter $\tau$ on the Galactic plane corresponding to the flat rotation curve in Fig. 2(a). The Galactic center is located at $(8.0,\,0)$ and the SSB is at $(0,\,0)$ point. The upper plot covers almost the whole Galactic plane, while the lower plot is the enlarged drawing for the vicinity of the Galactic center.}
\end{figure}
If we adopt the more realistic rotation curve as plotted in Fig. 2(b), the parameter within the radius $d_0$ can be written as
\begin{equation}
\tau = 2\pi\left(\frac{d_0}{r}\right)\left(\frac{c}{V_0}\right)~~~~~~~~(d<d_0),
\end{equation}
where Eq. (\ref{eq_vs}) has been taken into account. In this small region we have $r\simeq R_0 = 8.0~\rm{kpc}$, so that $\tau$ is approximately a constant:
\begin{equation}
\tau\simeq 320~~~~~~~~(d<d_0=0.3~\rm{kpc}),
\end{equation}
which indicates that during the light time $T_{\rm L}$ from the star to the observer, the star moves only about 1/320 of a circle in its orbit so that the curvature of the motion is not significant. Such small segment of arc can be considered as straight line and the acceleration of the star can be regarded as unchanged during its motion along the short arc. With an adoption of the rotation curve in Fig. 2(b), Eq. (\ref{eq_vec}) seems to be appropriate for all the stars on the Galactic disk.

\subsection{Aberration in proper motions for stars with short orbital periods}
We have shown in the last paragraphs that the Eq. (\ref{eq_vec}) is not appropriate to be used to describe the aberrational effect for the stars that have short orbital periods compared to the interval of light time from the star to the observer. According to the concept of the aberration in proper motions, after a time span covering integer multiple of orbital periods, the star moves back to the same point on the orbit which means that its velocity is the same as it is at the beginning of that time span, and the corresponding effect of aberration resulting from the acceleration of the star is zero. To this end, only orbital period fraction
\begin{eqnarray}
P_{\rm f} = {\rm remainder ~of}~ T_{\rm L} ~{\rm divided~ by}~ P
\end{eqnarray}
within the light time is responsible for effective aberration in proper motions, and Eq. (\ref{eq_vec}) should be written as a more suitable form:
\begin{eqnarray}
\label{eq_vec3}
\Delta\dot{ \bmath p}^{\rm S} = \frac{1}{c}\bmath p\times \left[\left(\bmath a^{\rm B} - \frac{\bmath v_2^{\rm S} - \bmath v_1^{\rm S}}{T_{\rm L}}\right) \times \bmath p\right],
\end{eqnarray}
where $\bmath v_1^{\rm S}$ and $\bmath v_2^{\rm S}$ are the velocities of the star on its orbit at the beginning ($t_1$) and the end ($t_2$) of the effective fraction of the period, respectively. The value of Eq. (\ref{eq_vec3}) also depends on the initial position of the star corresponding to the velocity $\bmath v_1^{\rm S}$ or the phase of the fraction (effective) arc if the orbit is not circular (e.g. elliptical orbit). To calculate the value for Eq. (\ref{eq_vec3}), it would require that the accuracy of the light time be measured to an accuracy at least better than $P_{\rm f}$.

Note that most of the short periodic stars are near the Galactic center, where the influence caused by the acceleration of the SSB can be ignored, we only consider the effect resulting from the motion of the star. Projecting Eq. (\ref{eq_vec3}) on the celestial sphere, we obtain the aberrational proper motions in the equatorial coordinate system:
\begin{equation}
\label{eq_pms20}
\Delta\mu_\alpha\cos\delta = \frac{1}{\kappa} \frac{v_{2,x}-v_{1,x}}{r}, ~~~
\Delta\mu_\delta = \frac{1}{\kappa} \frac{v_{2,y}-v_{1,y}}{r},
\end{equation}
where the subscript $x$, $y$ means that the velocity vectors are decomposed in the tangential coordinate system that is established by $(\bmath e_\alpha,\, \bmath e_\delta,\,\bmath e_r)$ triad as defined in Section 2.1, and $\kappa = 4.74047$ is a constant factor for unit transformation if proper motions are in unit of $\rm \mu as\,yr^{-1}$, velocities in $\rm km\,s^{-1}$, and $r$ in pc. Because the true proper motions of the star at $t_1$ are such that:
\begin{equation}
\left[\Delta\mu_\alpha\cos\delta\right]_1^{\rm true}  =  \frac{1}{\kappa} \frac{v_{1,x}}{r},~~~
\left[\Delta\mu_\delta\right]_1^{\rm true}  =  \frac{1}{\kappa} \frac{v_{1,y}}{r},
\end{equation}
we have the observed proper motions at $t_1$ by adding the corrections in Eq. (\ref{eq_pms20}):
\begin{eqnarray}
\left[\Delta\mu_\alpha\cos\delta\right]_1^{\rm obs} & = &\frac{1}{\kappa} \frac{v_{2,x}}{r} = \left[\Delta\mu_\alpha\cos\delta\right]_2^{\rm true} \nonumber \\
\left[\Delta\mu_\delta\right]_1^{\rm obs} & = &\frac{1}{\kappa} \frac{v_{2,y}}{r}
= \left[\Delta\mu_\delta\right]_2^{\rm true}.
\end{eqnarray}
This shows that the effect of aberration in proper motions only changes the observed phase of the star on the stellar orbit. It does not change the shape of the orbit on the celestial sphere, although we can not measure the exact value of the correction.

\section{Effect of the aberration on the celestial reference frame realized by Galactic stars in Gaia era}
The European space astrometry mission Gaia, due for launched in late 2013, will survey all the celestial objects (including stars, extragalactic sources, and solar system objects) brighter than 20th magnitude \citep{perryman01} with an unprecedent accuracy, ranging from a few tens of $\rm \mu as$ at magnitude  $15-18$ to about 200 $\rm \mu as$ at the the faint end \citep{lind08}. One target of the high accurate astrometric solution of Gaia is to build a kinematically non-rotating frame called Gaia-CRF (Gaia celestial reference frame) in optical and radio bands to a uniform precision \citep{mignard11}. It would be aligned with current ICRS realized by VLBI using certain optical bright extargalactic sources (\citealt{bourda08}; \citealt{bourda10}; \citealt{taris13}).

The systematic effect of the Galactic aberration on the present ICRF realized by extragalactic sources has been discussed by \citet{liu12} using ICRF1/ICRF2 catalogues. More recently, a new concept of `epoch ICRF' was proposed by \citet{xu13} with a slightly different approach. Because the extragalactic sources are assumed to have no measurable motion \citep{iers2010}, they can be regarded as free of acceleration, therefore only the acceleration of the SSB is necessarily be considered. For the reference system realized by stars in our Galaxy, the situation is more complicated. The fictitious proper motions are resulting from both accelerations as discussed in previous sections. The first part from the acceleration of the SSB causes a global rotation that depends on the distribution of the stars on the celestial sphere \citep{liu12}. On the other hand, the second part resulting from the stellar accelerations depends on the three dimensional places of the stars. Note that, while in \citet{liu12}, the global rotation of the ICRS due to the Galactic aberration effect is on the order of 1 $\rm \mu as~yr^{-1}$, the magnitude of global rotation derived from the aberration in proper motions for stars would be larger. The aberrational effect on the future stellar reference frame should be considered with micro-arcsecond astrometry, in order to construct a satisfactory inertial celestial reference system (at the level of about $0.5\,\rm \mu as\, yr^{-1}$, \citealt{mignard11}). However, the accelerations of stars, having an amplitude similar to that of the SSB's acceleration (several $\rm \mu as\, yr^{-1}$) are not easy to be detected with high accuracy one by one. Theoretical models such as more practical rotation curve are needed as proper simplification.

\section{Discussion and conclusion}
In this paper we have investigated the effect of aberration in proper motions for Galactic stars, which is an observational effect resulting from the accelerations of the observer and stars. This effect was first proposed by Kovalevsky in 2003, and will start to be noticeable in the era of micro-arcsecond astrometry with space missions.

The theoretical formulas of aberrational proper motions were developed with an hypothesis that the acceleration of the star is toward the Galactic center. In our study, two dimensionless parameters $\gamma$ (the ratio of the acceleration of the star and the SSB) and $\tau$ (ratio of the orbital period and the light time from the star to the SSB) were introduced. The parameter $\gamma$ specifies the significance of the aberration resulting from the star with respect to that from the observer, while $\tau$ determines whether the basic expression of the effect is applicable with sufficient appropriateness.

We have improved the results of \citet{kovalevsky03} to a more concise form. Two kinds of rotation curves of the Galactic disk were adopted to examine the property of the aberrational proper motions, especially in the vicinity of the Galactic center. A flat rotation curve starting from the Galactic center leads to enlargement of the proper motions to $1000\,\rm{\mu as\,yr^{-1}}$ at $d=0.2$ kpc, while the alternative rotation curve rising linearly from the Galactic center to the bulge-disk boundary gives limited proper motions up to about $150\,\rm{\mu as\, yr^{-1}}$. In this more practical case, the parameter $\tau$ is always higher than 300, which ensure the validity of the expression (\ref{eq_vec}) of the aberrational effect. If the period of the stellar orbit is shorter than the light time from the star to the observer, the assumption of constant acceleration in this period of time does not hold. This called attention to apply the aberration in proper motions with caution. In this circumstances, one need more basic expression as shown in Eq. (\ref{eq_vec3}). The magnitudes of the aberration in proper motions are difficult to measure, however we have shown that there is no effect on determining the orbit of stars.

Because the amplitudes of the systematic proper motions is at some places much larger than the Gaia accuracy for the proper motion measurements, this effect should be considered to eliminate the rotation and distortion in the future Gaia celestial reference system realized by stars in optical bandpass. However, this would be possible only if the accelerations of stars are known with satisfactory precision, or we have more reliable kinematics of the Galaxy for modeling those accelerations. Conventionally, proper motions of stars are derived from differentiate of positions (celestial coordinates in a specific reference system) at various epochs. The effect of aberration in proper motions is embedded in the observed positions of stars. In this sense, the systematic effect cannot be measured directly from observations; it can only be studied in a theoretical way. At the time of Gaia completion, the effect of aberration in proper motions is expected to be detected, e.g. using the least squares method as presented in \citet{titov11}, with the help of ultra high accurate observation data of the Gaia satellite.

\section*{acknowledgements}
The authors are grateful to Professor N. Capitaine (SYRTE, Paris Observatory), who provided very useful comments for improving the manuscript. This work is funded by the National Natural Science Foundation of China (NSFC) under grant No. 11173014.


\appendix
\section{An example of improper use of the expressions for the aberration in proper motions}
In the past decades, great progress has been made with high-resolution near-infrared observations of stars in the central cluster of our Galaxy\footnote{Generally, the kinematics of these stars does not obey the rule of rotation curves which are established in a statistical way.}. The experiments and analysis of proper motions and accelerations of stars near the Sgr A* have provide strong evidence for a massive black hole at the center of the Galaxy (\citealt{eckart96}; \citealt{ghez00}). The well determined orbital solutions for one typical star, S0-2, which was observed for almost the whole circle has been provided by \citet{schodel03} and the orbital elements are listed in Table A1. The acceleration of S0-2 is calculated from the Newton's law:
\begin{equation}
\bmath a^{\rm S0-2} = \frac{GM_{\rm BH}}{d^2}\bmath \rho,
\end{equation}
where $G$ is the gravitational constant, $M_{\rm BH}$ the enclosed mass of the central black hole.

\begin{table}
\caption{Elements of the Keplerian orbit of S0-2.}
\centering
\begin{tabular}{@{}lcr}
\hline
Element of the orbit & value & unit \\
\hline
Central mass & $3.3\pm 0.7$ & $10^6M_\odot$ \\
Semimajor axis & $4.54\pm0.27$ & $10^{-3}\,\rm pc$ \\
Separation of pericenter & $0.59\pm 0.10$ & $10^{-3}\,\rm pc$\\
Eccentricity & $0.87\pm0.02$ & $-$\\
Period & $15.73\pm 0.74$ & yr\\
Pericenter passage & $2002.31\pm0.02$ & yr \\
Inclination & $45.7\pm2.60$ & deg \\
Angle of line of nodes & $44.2\pm7.0$ & deg\\
Angle of node of pericenter & $244.7\pm4.7$ & deg \\

\hline
\end{tabular}

\medskip
The table is grabbed from Table 4 of \citet{schodel03}.
\end{table}

The coordinate system in which orbital elements are given is the local tangential coordinate system originated at Sgr A* in the equatorial coordinate system. This means that the three dimensional position $(\Delta x,\, \Delta y,\,\Delta z)$\footnote{$\Delta x,\, \Delta y,\,\Delta z$ are also the positional offsets of the star with respect to the Galactic center Sgr A*.} extracted from orbital elements are described in the coordinate system established by $(\bmath e_\alpha,\, \bmath e_\delta,\,\bmath e_r)$.

To calculate the proper motions in right ascension and declination using Eq. (\ref{eq_vec}), it is necessary to transform the position and velocity vectors into the equatorial coordinate system using the inverse relation of Eq. (\ref{eq_ead}) as follows:
\begin{eqnarray}
\label{eq_rho2}
\Delta X & = & +\Delta x\sin\alpha_0  +\Delta y\cos\alpha_0\sin\delta_0  - \Delta z\cos\alpha_0\cos\delta_0  \nonumber\\
\Delta Y & = & -\Delta x\cos\alpha_0  +\Delta y\sin\alpha_0\sin\delta_0  - \Delta z\sin\alpha_0\cos\delta_0  \\
\Delta Z & = & -\Delta y\cos\delta_0  - \Delta z\sin\delta_0;  \nonumber
\end{eqnarray}
in which $\alpha_0$ and $\delta_0$ are the equatorial coordinates of the compact radio source Sgr A* based on the best available VLBA observations \citep{reid04}:
\begin{eqnarray}
\label{eq_ad0}
\alpha_0 & = & 17^{\rm h}45^{\rm m}40\fs0400 \nonumber \\
\delta_0 & = & -29\degr00\arcmin28\farcs138.
\end{eqnarray}

\begin{figure}
\label{figa}
\centering
\includegraphics[width=80mm]{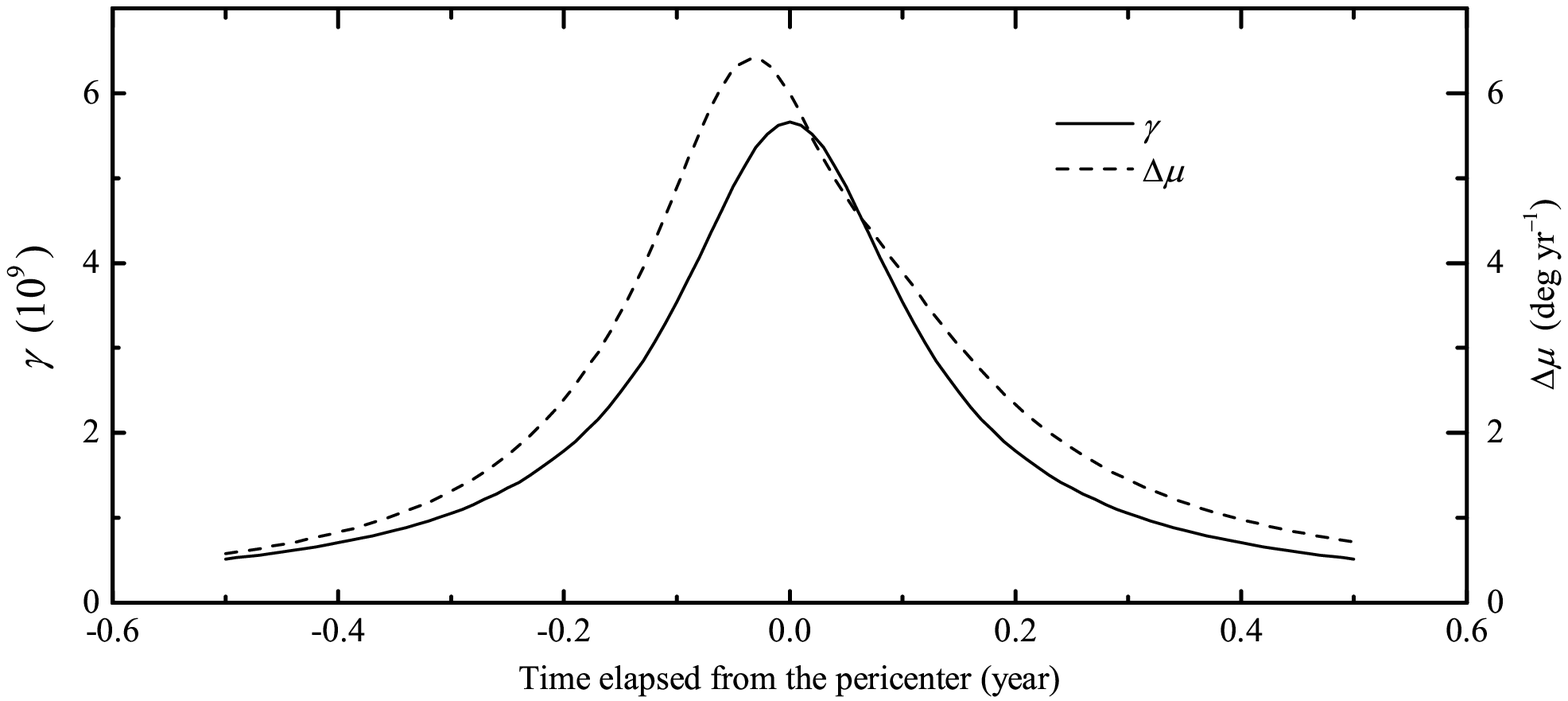}
\caption{The parameter $\gamma$ and the aberrational proper motions for the S0-2 star near the pericenter of the orbit. $\Delta \mu=\sqrt{\left(\Delta \mu_\alpha \cos \delta\right)^2+\Delta \mu_\delta^2}$ represents the general proper motion.}
\end{figure}

Because the observed data of stars are given in the equatorial coordinate system, we decompose the basic equation (\ref{eq_vec}) as follows:
\begin{eqnarray}
\label{eq_muad}
\Delta\mu_\alpha\cos\delta & = &\frac{1}{d} \gamma A^{\rm B}(
-\Delta X\sin\alpha+\Delta Y\cos\alpha)  \\
\Delta\mu_\delta & = &\frac{1}{d} \gamma A^{\rm B}(
-\Delta X\cos\alpha\sin \delta
-\Delta Y \sin\alpha \sin \delta
+ \Delta Z\cos \delta), \nonumber
\end{eqnarray}
where $d = \sqrt{\Delta x^2 + \Delta y^2 + \Delta z^2} = \sqrt{\Delta X^2 + \Delta Y^2 + \Delta Z^2}$, and $\alpha$, $\delta$ being the equatorial coordinates of S0-2:
\begin{eqnarray}
\label{eq_ad}
\alpha = \alpha_0 + \frac{\Delta x}{R_0};~~~~\delta = \delta_0 + \frac{\Delta y}{R_0}.
\end{eqnarray}
Inserting Eq. (\ref{eq_rho2}) into Eq. (\ref{eq_muad}), and taking into account Eq. (\ref{eq_ad}) we have the final expressions for the aberration in proper motions in the equatorial coordinate system:
\begin{eqnarray}
\label{eq_muad2}
\Delta\mu_\alpha\cos\delta  \simeq  \gamma A^{\rm B} \frac{\Delta x}{d}; ~~~~
\Delta\mu_\delta  \simeq  \gamma A^{\rm B} \frac{\Delta y}{d},
\end{eqnarray}
where the terms with higher order terms of $\Delta x$ and $\Delta y$ are neglected. Alternatively, the aberrational proper motions can be derived from the geometrical concept of the aberrational effect as stated in Sect. 3. The light time from the S0-2 to the observer is $T_{\rm L} = R_0/c$, and the acceleration projected to the directions of right ascension and declination are $a^{\rm S}\cdot \Delta x/d$ and $a^{\rm S}\cdot \Delta y/d$, respectively. Then the angular proper motion during the light time from the star to the observer are calculated as if the acceleration is a constant:
\begin{eqnarray}
\Delta\mu_\alpha\cos\delta  \simeq  a^{\rm S}T_{\rm L} \frac{\Delta x}{d} \frac{1}{R_0}, ~~~~
\Delta\mu_\delta  \simeq a^{\rm S}T_{\rm L} \frac{\Delta y}{d} \frac{1}{R_0}.
\end{eqnarray}
Clearly, they are consistent with the results in Eq. (\ref{eq_muad2}).

Shown in Fig. A1, the parameter $\gamma$ for S0-2 is on the order of $10^7-10^9$, and the proper motions can be up to several degrees per year. This appears unrealistic because the orbital period of S0-2 is only about  $15\,{\rm yr}$, while the light time from S0-2 to the SSB is about $T_{\rm L} \simeq 8000\,{\rm pc} \times 3.26~{\rm yr\,pc^{-1}} = 26000\,\rm yr$, which is about 1700 times larger than $P$ ($\tau\simeq1/1700$). The calculation using Eq. (\ref{eq_vec}) extrapolates the acceleration at the staring point to the whole time span as if the acceleration was a constant, and this inappropriate procedure causes accumulated high proper motion corrections.

\bsp

\label{lastpage}

\end{document}